\documentstyle[prl,aps,amsfonts,epsfig]{revtex}
\begin{document}
\draft
\title{Spatio-temporal structure of traffic flow in a system with 
an open boundary}
\author{Namiko Mitarai and Hiizu Nakanishi}
\address{Department of Physics, Kyushu University 33, Fukuoka 812-8581, Japan}
\date{March 16, 2000}

\maketitle
\begin{abstract}
The spatio-temporal structure of traffic flow pattern is investigated
under the open boundary condition using the optimal velocity (OV)
model. The parameter region where the uniform solution is convectively
unstable is determined. It is found that a localized perturbation triggers a
linearly unstable oscillatory solution out of the linearly unstable
uniform state, and it is shown that the oscillatory solution is also
convectively stabilized. It is demonstrated that the observed traffic
pattern near an on-ramp can be interpreted as the noise sustained
structure in the open flow system.
\end{abstract}
The complex behavior of traffic flow on a freeway has
been attracting the research interests \cite{KR,K}.  
In recent years, various kind
of dynamical states have been observed in real traffic; a peculiar
fluctuating flow called {\it the synchronized flow} (SF) was found in the
high density region which could result in the jammed
flow \cite{KR,K}.  
The flux of cars is higher than the jammed flow, but
fluctuates synchronously between different lanes.  It has been found
that SF is triggered by a localized perturbation such as an on-ramp,
and may persist for several hours. 
{\it The stop-and-go state} (SGS) is another state of traffic where the
traffic goes through an alternating pattern of jammed and free flows in
a short period of space and time; 
SGS is often observed behind SF region \cite{K}.  
The hysteresis has been also found in the transition
between the jammed flow or SF and the free flow when the car density
changes \cite{KR,K}.

One of the major topics in the current theoretical research on the
traffic flow is to make clear the characteristics of the complex
behavior \cite{Hmodels,MiN99,TSH00,CA2}.  
Recently, the effects of the localized perturbation were
investigated in the hydrodynamical traffic flow models with an on-ramp
under the open boundary condition\cite{Hmodels}.  
It has been found that the influx
perturbation at the on-ramp causes various types of oscillatory flows
as well as the convectively stabilized uniform dense flow, but
properties of these oscillatory flows and spatio-temporal patterns
have not been understood yet.

The optimal velocity (OV) model is another type of model based on the
driving behavior of individual cars \cite{OVM}.  The OV model has been
demonstrated to show the transition from free flow to jammed flow,
and to reproduce the flux-density diagram (the fundamental
diagram) similar to the one observed in the real traffic \cite{OVM2}.
The authors have examined the effects of a localized perturbation 
in the OV model and found that the localized perturbation
triggers the oscillatory flow out of the linearly unstable uniform flow
\cite{MiN99}.

In the present work, we study the convective instability in the OV model
and clarify
the mechanism of the spatio-temporal pattern triggered by a localized
perturbation.  We show the resulting spatio-temporal pattern is 
analogous to the one observed in the real traffic and can be
understood using the idea of the noise sustained structure
in the open flow system \cite{D89}.

The OV model \cite{OVM} is the one-dimensional car-following model
where each driver tends to drive 
at the optimal velocity determined by the headway of his car.
The position of the $n$th car $x_n(t)$ at time $t$
obeys the equation of motion
\begin{equation}
   \ddot x_n(t)=a\left[ U(b_n(t))-\dot x_n(t)\right] ;
   \quad
   b_n(t)\equiv x_{n+1}-x_{n},
\label{eq:OVM}
\end{equation}
where the dots denote the time derivative and $b_n$ represents the
headway of the $n$th car; we assume the $(n+1)$th car precedes the $n$th
car.  The parameter $a$ is a sensitivity constant, and the function
$U(b)$ determines the optimal velocity for a driver when his headway is
$b$.  For $U(b)$, we employ
\begin{equation}
U(b)=\tanh (b-2)+\tanh(2),
\end{equation}
as other works on the OV model \cite{OVM,KS95}.

Equation (\ref{eq:OVM}) has a uniform solution
\begin{equation}
x_n(t)=\bar{b}n+U(\bar{b})t,
\label{eq:uniform}
\end{equation}
where all the cars go with the same headway $\bar b$
and the same speed $U(\bar b)$.
The dispersion for the small deviation around the uniform solution in
the ``index frame'' is given by
\begin{equation}
   \omega =
       -i\frac{a}{2}+\frac{i}{2}\sqrt{a^2+4aU'(\bar b)(e^{i k}-1)}
   \equiv \omega_I( k)
\end{equation}
where $\omega$ and $k$ denote the angular frequency and the 
wave number in the index frame; $\exp(i kn-i\omega t)$.
From this, it can be seen that the uniform solution is linearly unstable 
when
\begin{equation}
a<2U'(\bar{b}),
\label{eq:unstable}
\end{equation}
where the prime denotes the derivative by its argument \cite{OVM}.
When we perturb the linearly unstable uniform solution 
under the periodic boundary condition, 
the effect of perturbation grows in time
and eventually the system segregates into two regions;
the jammed region with smaller headway and lower velocity,
and the free flow region with larger headway and higher
velocity \cite{OVM}.
It is also known that Eq.\ (\ref{eq:OVM}) is reduced to the Korteweg-de
Vries (KdV) or the modified KdV equation in weak nonlinear analysis
near the linear stability limit \cite{KS95}

We now study the system behavior in the situation where the upper and
lower stream are distinguished, which is more appropriate to the actual
traffic.
We employ the open boundary condition defined as follows:
(i)At the upper stream end ($x=0$), 
cars with the velocity $U(\bar b)$ enter the system with 
the constant time interval $\bar b/U(\bar b)$.
(ii)Around the lower stream end,
the car that is farthest ahead, which has no car to follow within
the system, obeys the equation of motion
$\ddot{x}_{f}=a\left[U(\bar{b})-\dot{x}_{f}\right]$,
and it goes out of the system at $x=L$.
Here, $\bar b$ is chosen to fit the uniform initial state, which will go on
if there is no perturbation.

The uniform solution is perturbed locally in
space and time by shifting the velocity of the $0$th car at $t=0$
by a small value $\epsilon$.
In actual simulations, the initial condition is given as:
\begin{equation}
x_n(0) = \bar b n+\frac{L}{2},\
\dot x_n(0) = U(\bar b) \
\mbox{ for } n = \pm 1, \pm 2, ...,
\end{equation}
\begin{equation}
x_0(0) = \frac{L}{2},
\
\dot x_0(0) = U(\bar b)+\epsilon.
\end{equation}

\begin{figure}
\begin{center}
\epsfig{file=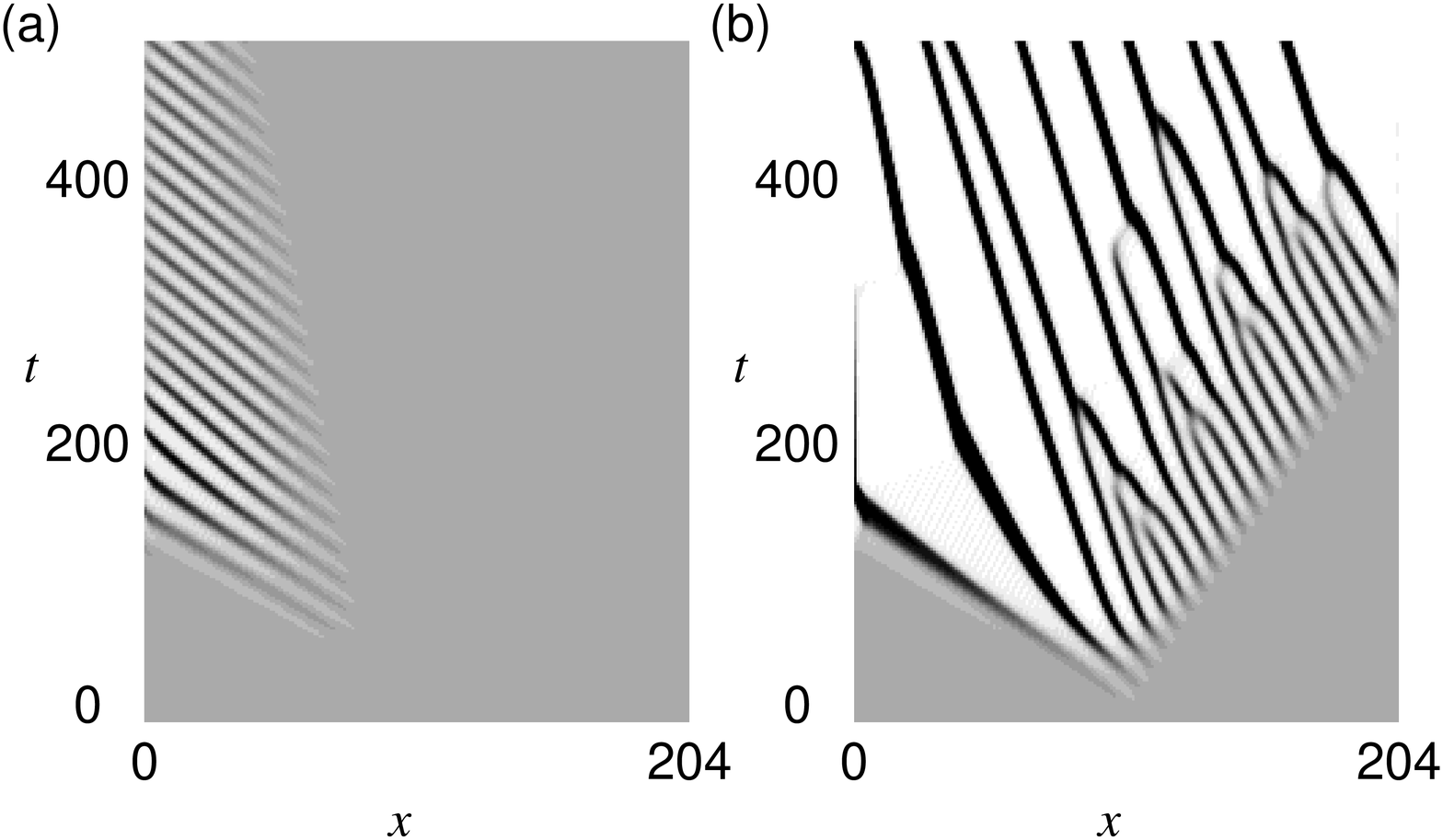,width=8.5cm}
\end{center}
\caption{The spatio-temporal diagrams of density. 
The horizontal axis is the position of a car
$x$ and the vertical axis is the time $t$.
The higher density region is shown by a darker region. 
The darkness is adjusted so that the initial uniform flow 
region is shown by a gray region.
(a)$a=1.4$, $\bar b=2.0$, $\epsilon=0.1$, and $L=204$.
The disturbed region is convected only upstream.
(b)$a=1.0$, $\bar b=2.0$, $\epsilon=0.1$, and $L=204$.
The disturbed region spreads in both directions.
}
\label{ovfig}
\end{figure}

Within the parameter region where the initial uniform solution is
linearly unstable, there are following two regions: When $a$ is larger
than a critical value $a_c$ which depends on $\bar b$ ($a_c=a_c(\bar
b)<a<2U'(\bar b)$), the disturbance travels only upstream (Fig.\
\ref{ovfig}(a)). Therefore, the disturbed region eventually goes out
from the system, and the linearly unstable uniform solution is
recovered.  On the other hand, when $a<a_c(\bar b)$, the disturbance
travels in both directions (Fig.\ \ref{ovfig}(b)), and the uniform flow
region is eliminated completely.  In the former case, the uniform
solution is {\it convectively unstable}; the growing perturbation is
convected away from any fixed location.  The uniform solution is {\it
absolutely unstable} in the latter case; the perturbation grows at every
point in space \cite{D89,LP81}.  The boundary between the convective
instability and the absolute one depends on the reference frame.  It
should be noted that, for a car-following model where one does not feel
any effects from behind as in the OV model, the instability of the
linearly unstable uniform solution is always convective in the index
frame, that moves with the cars.

The stability in the {\it laboratory frame} can be examined by 
following the procedure described in Ref. \cite{LP81}.
The dispersion
relation in the laboratory frame is given by
\begin{equation}
\omega(k) = k{U(\bar b)\over \bar b} + \omega_I(k)
\end{equation}
because the laboratory frame is moving with the speed $-U(\bar b)/\bar b$
relative to the index frame.
The convective stability limit $a_c(\bar b)$ is determined for a given
$\bar b$ by the set of equations;
\begin{equation}
\left.\frac{d\omega (k)}{dk}\right|_{k=k_c}=0,
\quad
\Im[\omega(k_c)]=0.
\label{bound:conv}
\end{equation}

\begin{figure}
\begin{center}
\epsfig{file=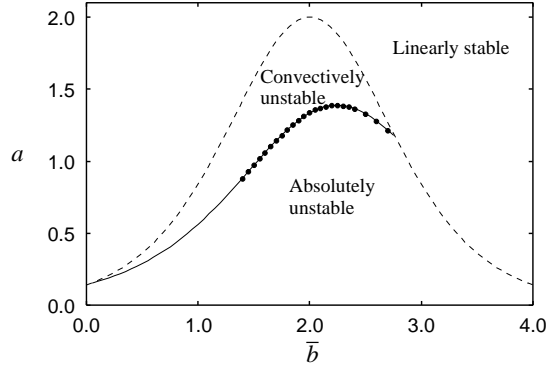,width=8.5cm}
\end{center}
\caption{The parameter region where the uniform solution is convectively 
unstable.
The solid line is the parameter boundary $a=a_c(\bar b)$ and
the dashed line is the linear stability limit $a=2U'(\bar b)$.
The uniform solution is convectively unstable in the region
between the solid line and the dashed line.
The open circles are the parameter $a=a_c(\bar b)$
estimated by the numerical simulations.}
\label{convecfig}
\end{figure}

Fig.\ \ref{convecfig} shows numerical estimate of Eq.\ (\ref{bound:conv})
for the boundary $a=a_c(\bar b)$ (the solid line) with the linear
stability limit $a=2U'(\bar b)$ (the dashed line). The uniform
solution is convectively unstable in the region between the solid line
and the dashed line ($a_c(\bar b)<a<2U'(\bar b)$).


Another characteristic of the density diagram is
its spatio-temporal pattern; the oscillatory flow (a regular stripe 
in Fig.\ \ref{ovfig}(b)) followed by
an alternating sequence of jams and free flows 
(an irregular stripe with stronger contrast in Fig.\ \ref{ovfig}(b)).
This structure is triggered out of the linearly unstable uniform
solution by the localized perturbation.

\begin{figure}[htb]
\begin{center}
\epsfig{file=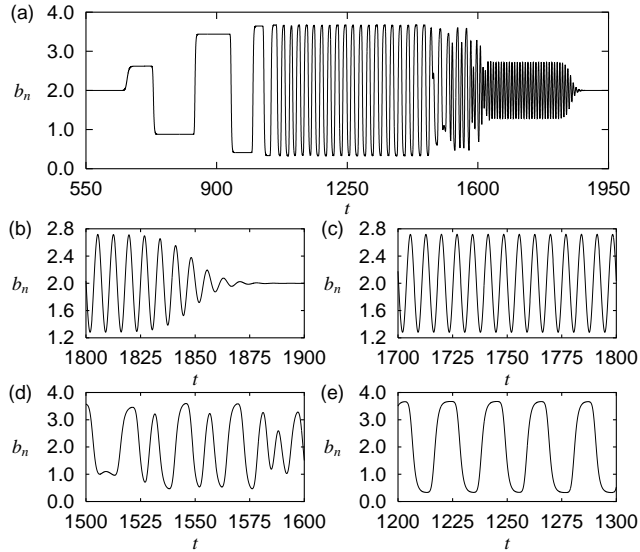,width=8.5cm}
\end{center}
\caption{The time evolution of the $n=-578$th car's headway $b_n(t)$ with
$a=1$, $\bar b=2$, $\epsilon=0.1$, and $L=800$.
(a)The effect of the perturbation has not reached to
the both ends of the system yet. Thus, the linearly unstable uniform flow 
regions are left and the whole structure can be seen.
(b)The downstream front of the disturbed region.
(c)The oscillatory flow region.
(d)The oscillatory flow breaks up.
(e)The alternating sequence of jams and free flows, which is 
not completely periodic.}
\label{patternfig}
\end{figure}

This sequence can be seen more clearly in Fig.\ \ref{patternfig}, in
which the time evolution of the headway of the $-578$th car is shown.
Each car travels through the sequence, thus the time evolution of the
headway shows the structure of flow from the upper to the lower stream.
In Fig.\ \ref{patternfig}(a), the car passes through the alternating
jammed and free flow region during $1000\lesssim t\lesssim 1400$, and
the oscillatory flow region during $1600\lesssim t\lesssim 1800$.  We
can see the periodic behavior of the headway in the oscillatory flow
region (Fig.\ \ref{patternfig}(c)).

This oscillatory behavior of the solution can be expressed by
\begin{equation}
b_n=\hat{b}+f(n-ct)
\label{eq:oscillate}
\end{equation}
with an appropriate phase speed $c$ and a function $f$, which, we
assume, has zero mean by taking $\hat b$ as the mean headway. We have
already found that Eq.\ (\ref{eq:oscillate}) can be a solution of the
original equation of motion (\ref{eq:OVM}) for a finite range of the
phase speed $c$ \cite{MiN99}; e.g., $c\in (-0.637,-0.556)$ for $a=1.0$
and $\hat b=2.0$.  The shape, the wavelength, and the amplitude of the
solution (\ref{eq:oscillate}) depend upon the parameters $a$, $\hat b$,
and $c$. When we set the phase speed at $c_s$, the value
obtained from the direct simulation of Eq.\ (\ref{eq:OVM}), namely,
$c=c_s=-0.610$ for $a=1.0$ and $\bar b=2.0$, we get $f(n-c_st)$ which
coincides with the results of the simulation (see Fig.\ 5(c) in Ref.\
\cite{MiN99}).

The linear stability of periodic solutions can be determined by the
Floquet exponent, or a complex linear growth rate averaged over the
period \cite{H83}.  We calculate the Floquet exponents of the
oscillatory solution for the N-car system with the periodic boundary
condition, where the headway of the $N+1$th car equals to that of the
first car, and the fixed boundary condition, where the headway of the
$N+1$th car obeys the oscillatory solution.  It is found that the
maximum value of the real part of the exponents is always
positive under the periodic boundary condition, while it is always
negative under the fixed boundary condition.
%
%
This implies the oscillatory solution is
linearly unstable, 
but the growing disturbance is convected away if the headway of the
foremost car obeys the oscillatory solution; namely, the oscillatory
solution is convectively unstable in the index frame.

We now analyze the mechanism how the solution with a particular phase speed
is chosen out of the finite range of allowed ones.  Suppose
the downstream front of the oscillatory region is propagating with the
speed $V_0$ relative to the index frame, then $V_0$ and the wave number
$k_f$ in the index frame can be determined by the condition that the
linear growth rate in the frame moving with the front is zero;
\begin{equation}
\left.{d \omega_{V_0}(k)\over dk}\right|_{k=k_f}=0,
\quad
\Im [ \omega_{V_0}(k_f)]=0,
\end{equation}
where $\omega_{V_0}(k)\equiv\omega_I(k)-kV_0$ is the dispersion in
the moving frame.  The angular frequency of the oscillation at the
front in the moving frame is given by $\Re[\omega_{V_0}(k_f)]$, and this
oscillation is amplified as it travels upstream to induce the
oscillatory solution.  It is natural to expect that the time
period of the oscillatory solution is the same as that of the downstream
oscillation in the moving frame because each car tends to follow the
motion of the preceding car (see Eq.\ (\ref{eq:OVM})).
Therefore, there should be the relationship between the wave length
$\lambda$ of the oscillatory solution and the phase speed $c_s$,
\begin{equation}
\lambda=(|c_s|+V_0)\frac{2\pi}{|\Re[\omega_{V_0}(k_f)]|}.
\label{eq:lambda}
\end{equation}
This was confirmed by the simulations as is shown in Table \ref{lambda}.

This oscillatory flow cannot extend over the whole system because it is
only convectively stable.  The motions of cars in the upper stream
gradually deviate from the oscillatory solution, and eventually the
oscillatory flow breaks up (Fig.\ \ref{patternfig}(d)).  As a result,
the alternating sequence of jams and free flows is formed
behind the oscillatory flow region.  The alternating region cannot be
completely periodic because any infinitesimal perturbation grows as it
travels upstream (Fig.\ \ref{patternfig}(e)).  This mechanism of the
structure formation is general to 
convectively unstable open flow systems, such as the complex
Ginzburg-Landau equation with an advection term \cite{D89}.


Before concluding, we discuss the present results in connection with
the other traffic models and the observation of real traffic.
%
As we have seen, the convective instability in the index frame plays
important role in the structure formation.  It should be a common
feature of car-following models
and the main feature of the
flow pattern obtained should not depend on the details of the 
models \cite{comment}.

The hydrodynamical models based on the continuum description have
been found to show the similar behavior to the 
oscillatory flow \cite{Hmodels}.
The oscillatory flows appear near an on-ramp with influx in the
simulations under the open boundary condition \cite{comment2}.
It has been also found that the convectively unstable uniform flow can
be realized in the upper stream of the on-ramp in some parameter
region.  Therefore, it is natural to expect that a {\it noise-sustained
structure} \cite{D89} similar to the one discussed in the present work
appears in the hydrodynamical models when a small noise perturbs the
convectively unstable uniform flow.

In the cellular automata (CA) models, there should be an analogous
phenomenon to the convective instability although the idea of linear
stability does not exist; the effect of localized perturbation travels
only backward in the frame moving with cars when car-car interaction
usually determined by gap in front of a car \cite{CA2,CA}.  On the other
hand, the oscillatory flow is more difficult to realize because of the
discreteness of the dynamical variables.  Recently, however, multi-value
extension of CA models was considered and it was reported that the flow
pattern similar to SGS can be realized by perturbing a meta-stable
uniform state\cite{CA2}.  In such a CA model, the mechanism that causes
the complex flow pattern discussed here may hold.

In the real traffic, it has been observed near an on-ramp that SF is
followed by SGS towards the upper stream, i.e. each car experiences the
fluctuating high flux flow after going through alternate jammed and free
flow regions as approaching the on-ramp \cite{K}. This can be interpreted as
follows: First, the convectively unstable uniform flow region is formed
near the on-ramp by the influx.  Then, small noise in the flow induces
the convectively stabilized oscillatory flow, which corresponds to SF.
The oscillatory flow breaks up as it travels towards the upper stream,
and many small jams are formed in the upper stream side of the
oscillatory flow region, which is SGS.  This pattern is maintained
near the on-ramp by small noise, namely, the noise-sustained structure.

Summarizing our results, we have studied the convective instability of the
uniform flow solution in the OV model and shown that a localized
perturbation to it generates the sequence of flow patterns; the
oscillatory flow followed by the alternating sequence of jams and free
flows towards the upper stream.  We have demonstrated that the
oscillatory solution is linearly unstable but is stabilized
convectively, and have clarified the selection mechanism of the
oscillatory solution out of the possible range of wave length.  It is
shown that the real traffic flow pattern observed near an on-ramp can be
interpreted using the idea of noise-sustained structure in the open flow 
system.

N.M. is grateful to K. Fujimoto for the informative discussions.

\begin{table}[h]
\caption{
The values of $\lambda$ from 
Eq.\ (\ref{eq:lambda}) and from the results of the numerical simulation.
}
\begin{tabular}{cccccc}
$\bar b$&$a$&$\lambda$ (Eq.\ (\ref{eq:lambda}))&$\lambda$
 (simulation)
\\ \hline
2.0 &1.0 &4.35&4.36\\
2.0 &1.5 &6.31&6.35\\
2.2&$2U'(\bar b)-1.0$ &4.30&4.30\\
2.2&$2U'(\bar b)-0.5$ &6.23&6.28\\
1.8&$2U'(\bar b)-0.5$ &6.23&6.28\\ 
\end{tabular}
\label{lambda}
\end{table}


\begin{references}
\bibitem{KR}
B. S. Kerner and H. Rehborn,
Phys. Rev. E {\bf 53}, R4275 (1996);
Phys. Rev. Lett. {\bf 79}, 4030 (1997).

\bibitem{K}
B. S. Kerner,
Phys. Rev. Lett. {\bf 81}, 3797 (1998).

\bibitem{Hmodels}
D. Helbing, A. Hennecke, and M. Treiber,
Phys. Rev. Lett. {\bf 82}, 4360 (1999);
H. Y. Lee, H. W. Lee, and D. Kim,
Phys. Rev. E {\bf 59}, 5101 (1999).

\bibitem{MiN99}
N. Mitarai and H. Nakanishi,
J. Phys. Soc. Jpn. {\bf 68}, 2475 (1999).

\bibitem{TSH00}
E. Tomer, L. Safonov, and S. Havlin,
Phys. Rev. Lett. {\bf 84}, 382 (2000).

\bibitem{CA2}
K. Nishinari and D. Takahashi,
Preprint, nlin/0002007.

\bibitem{OVM}
M. Bando, K. Hasebe, A. Nakayama, A. Shibata, and Y. Sugiyama,
Jpn. J. Ind. Appl. Math. {\bf 11}, 203 (1994);
Phys. Rev. E {\bf 51}, 1035 (1995).

\bibitem{OVM2}
M. Bando, K. Hasebe, K. Nakanishi, A. Nakayama, A. Shibata, and Y. Sugiyama,
J. Phys. I France {\bf 5}, 1389 (1995).

\bibitem{D89}
R. J. Deissler,
J. Stat. Phys. {\bf 54}, 1459 (1989).

\bibitem{KS95}
T. S. Komatsu and S. Sasa, Phys. Rev. E {\bf 52}, 5574 (1995).

\bibitem{LP81}
E. M. Lifshitz and L. P. Pitaevskii,
 {\it Physical Kinetics} 
(Pergamon, Oxford, 1981).

\bibitem{H83}
H. Haken,
{\it Advanced synagetics: Instability hierarchies of self-organizing
  systems and devices
}
(Springer-Verlag, Berlin, 1983).

\bibitem{CA}
K. Nagel, Phys. Rev. E  {\bf 53}, 4655 (1996).

\bibitem{comment}
The possibility of the pattern formation in a car-following model
has been already pointed out in Ref. \cite{D89}.

\bibitem{comment2}
It is not clear that these two types of oscillatory flows have 
the same physical origin because, in the present model, only a few
cars are involved within a period of the oscillation, therefore,
the validity of the continuum description is not obvious.

\end{references}
\end{document}